\documentstyle[twoside,fleqn,fps,espcrc2]{article}

\newcommand{\AmS}{{\protect\the\textfont2
  A\kern-.1667em\lower.5ex\hbox{M}\kern-.125emS}}

\def\lsi{\raise0.3ex\hbox{$<$\kern-0.75em\raise-1.1ex\hbox{$\sim$}}}
\def\gsi{\raise0.3ex\hbox{$>$\kern-0.75em\raise-1.1ex\hbox{$\sim$}}}

\newcommand{\gsim}{\mathop{\gsi}}
\newcommand{\R}{{\kern+.25em\sf{R}\kern-.78em\sf{I} 
  \kern+.78em\kern-.25em}}
\newcommand{\C}{{\kern+.25em\sf{C}\kern-.50em\sf{I} \kern+.50em\kern-.25em}}
\newcommand{\eps}{\epsilon}
\newcommand{\be}{\begin{equation}}
\newcommand{\ee}{\end{equation}}
\newcommand{\bea}{\begin{eqnarray}}
\newcommand{\eea}{\end{eqnarray}}
\newcommand{\nn}{\nonumber}
\newcommand{\la}{\langle}
\newcommand{\ra}{\rangle}
\newcommand{\vnu}{\vert \nu \vert}

\title{Simulations
in the $\eps$-Regime of Chiral Perturbation Theory
\thanks{Poster presented by K.-I. Nagai at Lattice 2003
\newline \hspace*{0.5mm} Preprint DESY 03-126, HU-EP-03/56, SFB/CPP-03-31
}}
\author{K.-I. Nagai
\address{ NIC/DESY Zeuthen, Platanenallee 6, D-15738 Zeuthen, Germany \\
$^{{\rm b}}$ Institut f\"{u}r Physik, Humboldt Universit\"{a}t zu Berlin,
Newtonstr. 15, D-12489 Berlin, Germany }
, W. Bietenholz $^{{\rm b}}$, T. Chiarappa $^{{\rm a}}$,
K. Jansen $^{{\rm a}}$ and S. Shcheredin $^{{\rm b}}$  
}

\begin{document}

\begin{abstract}

We discuss the potential of Ginsparg-Wilson fermion simulations in the
$\epsilon$-regime of chiral perturbation theory, regarding the
determination of the leading low energy constants of the effective
chiral Lagrangian. 
It turns out to be very hard to measure observables in the
topologically trivial sector. There a huge statistics would be required,
due to the frequent occurrence of very small eigenvalues.
Moreover, contact with chiral perturbation theory is only established
if the physical volume of the system is sufficiently large.

\vspace*{-4mm}

\end{abstract}

\maketitle

QCD at low energy can be described by an effective Lagrangian 
${\cal L}_{\rm eff}$ for the quasi-Goldstone bosons of chiral
symmetry breaking. ${\cal L}_{\rm eff}$ may then be evaluated by
chiral perturbation theory ($\chi$PT).
The case of a finite volume and very light 
quarks --- so that the pion Compton wave length clearly exceeds the box 
length --- is called the {\em $\eps$-regime}. There the zero-mode is 
treated by collective variables, 
while the excitations (which do fit into the box)
can be evaluated perturbatively in the $\eps$-expansion \cite{GasLeu}.
In this framework, observables strongly depend on the topology 
of the gauge background \cite{LeuSmi}, hence they should be measured in
fixed topological sectors.

${\cal L}_{\rm eff}$ involves coupling constants which appear as free
parameters in $\chi$PT. In principle they can be determined by 
comparison with numerical data from lattice QCD in the $\eps$-regime. 
The same values are also relevant for
the physical situation in a large volume (the standard $\chi$PT), 
hence their knowledge is
very important in view of experiments and further numerical studies,
especially for the chiral extrapolation.
Their numerical evaluation, however, may be easier in the
$\epsilon$-regime, due to the usage of a small volume.

Such simulations in the $\eps$-regime are now conceivable by using
Ginsparg-Wilson fermions, which avoid additive mass renormalization and
define the topological charge through the Index Theorem \cite{Has}.
Here we discuss the question how one could extract
results for the low energy constants
which occur in the leading order of ${\cal L}_{\rm eff}$,
the pion decay constant $F_{\pi}$ and the scalar condensate
$\Sigma$.

We performed quenched QCD simulations with the Wilson gauge action 
and the overlap Dirac operator \cite{Neu}
\bea
\hspace*{-6mm} &&
D_{\rm ov} = \Big( 1 - \frac{m}{2\mu} \Big) D_{ov}^{(0)} + m \ , \\
\hspace*{-6mm} &&
D_{\rm ov}^{(0)} = \mu (1 + A /\sqrt {A^{\dagger}A})\ , 
\quad A = D_W - \mu \ , \nn
\eea
where $m$ is the quark mass and $D_W$ the Wilson operator.
At $\beta =6$ and $\beta =5.85$ we set $\mu = 1.4$ resp.\ $\mu = 1.6$.
We approximated the inverse square root by Chebyshev polynomials
to an accuracy of $10^{-12}$.
To identify the low lying eigenvalues (EVs) of 
$D^{(0)~\dagger}_{\rm ov} D_{\rm ov}^{(0)}$ we applied alternatively the 
Ritz functional method and the Arnoldi algorithm. The latter
allows for the determination of many
EVs ($O(100)$). The very lowest EVs determine
the index $\nu$: there is usually a gap by several orders of magnitude
between the numerical values of the zero EVs and the rest of the spectrum.
The chirality of the zero modes 
reveals the sign of $\nu$. 
Two typical examples are given in Table \ref{EVsample}.

\begin{table}
\begin{flushright}
\begin{tabular}{|c|c|}
\hline
EVs of $D^{(0)~\dagger}_{ov} D_{ov}^{(0)}$ & chirality \\
\hline
\multicolumn{2}{|c|}{example for charge $\nu =0$} \\
\hline
$6.23013 e-3$  & $~~0.3608201$ \\
$6.23014 e-3$  & $-0.3608198$ \\
$9.90174 e-3$  & $-0.7622260$ \\
$9.90177 e-3$  & $~~0.7622260$ \\
$2.69086 e-2$  & $-0.7528350$ \\
\hline
\multicolumn{2}{|c|}{example for charge $\nu = -2$} \\
\hline
$6.03761 e-16$ & $-1.0000000$ \\
$9.43764 e-11$ & $-0.9999999$ \\
$3.68203 e-3$  & $-0.9240370$ \\
$3.73623 e-3$  & $~~0.9205659$ \\
$7.04176 e-3$  & $-0.7016518$ \\
\hline
\end{tabular}
\end{flushright}
\vspace*{-4mm}
\caption{\it{Typical examples for the lowest eigenvalues
of $D^{(0)~\dagger}_{\rm ov} D_{\rm ov}^{(0)}$ on a $10^3 \times 24$ lattice
at $\beta =6$.}}
\label{EVsample}
\vspace*{-9mm}
\end{table}

On a $10^3 \times 24$ lattice at $\beta = 6$ we also compared the index
with the topological charge determined from
(standard) cooling. In this respect, we investigated 51 configurations: 
for 41 of them the index of $D_{\rm ov}$ and the charge determined from 
cooling coincided, and in the remaining 10 cases it differed by 1. 

\begin{figure}[hbt]
\vspace*{-2mm}
\def\fpsangle{270}
\epsfxsize=50mm
\fpsbox{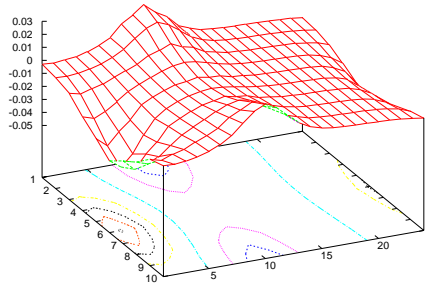}
\vspace*{-12mm}
\caption{\it{A 2-d cut of the configuration discussed in the text after
200 cooling sweeps. The vertical axis shows the naive topological charge
density, which extends over both signs, suggesting something
like an instanton anti-instanton pair.}}
\label{IAI}
\vspace*{-8mm}
\end{figure}

Next we mention that Random Matrix Theory (RMT) has been applied to
QCD and it yields predictions for the probability distributions
of the low lying eigenvalues \cite{RMT}. For the first non-zero
EV these predictions are confirmed if the physical
volume is sufficiently large, $V \gsim (1.2~{\rm fm})^{4}$ \cite{BJS}.
In particular, we know from these probability distributions that
in the topologically trivial sector there is a significant density
of {\em very small} (non-zero) EVs.
As an example, we found a configuration on the
$10^3 \times 24$ lattice at $\beta =6$ where the lowest EVs
of $D^{(0)~\dagger}_{\rm ov} D_{\rm ov}^{(0)}$ read:
$7.32e-5$, \ $7.33e-5$, \ $1.04e-4$ etc. The corresponding
chiralities $-0.127$, \ $0.127$, \ $0.826$ show clearly
that this configuration has index 0. This is also confirmed from cooling,
but the form of the cooled configuration suggests that charge
0 arises from the cancellation of a pair of topological objects with
charges $\pm 1$, see Fig.\ \ref{IAI}. The index never involved such a
cancellation. However, in the case of small non-zero EVs
there seems to be a trend towards a cancellation picture from cooling,
as in the example discussed here.

\begin{figure}[hbt]
\begin{tabular}{cc}
\def\fpsangle{0}
\epsfxsize=44mm
\fpsbox{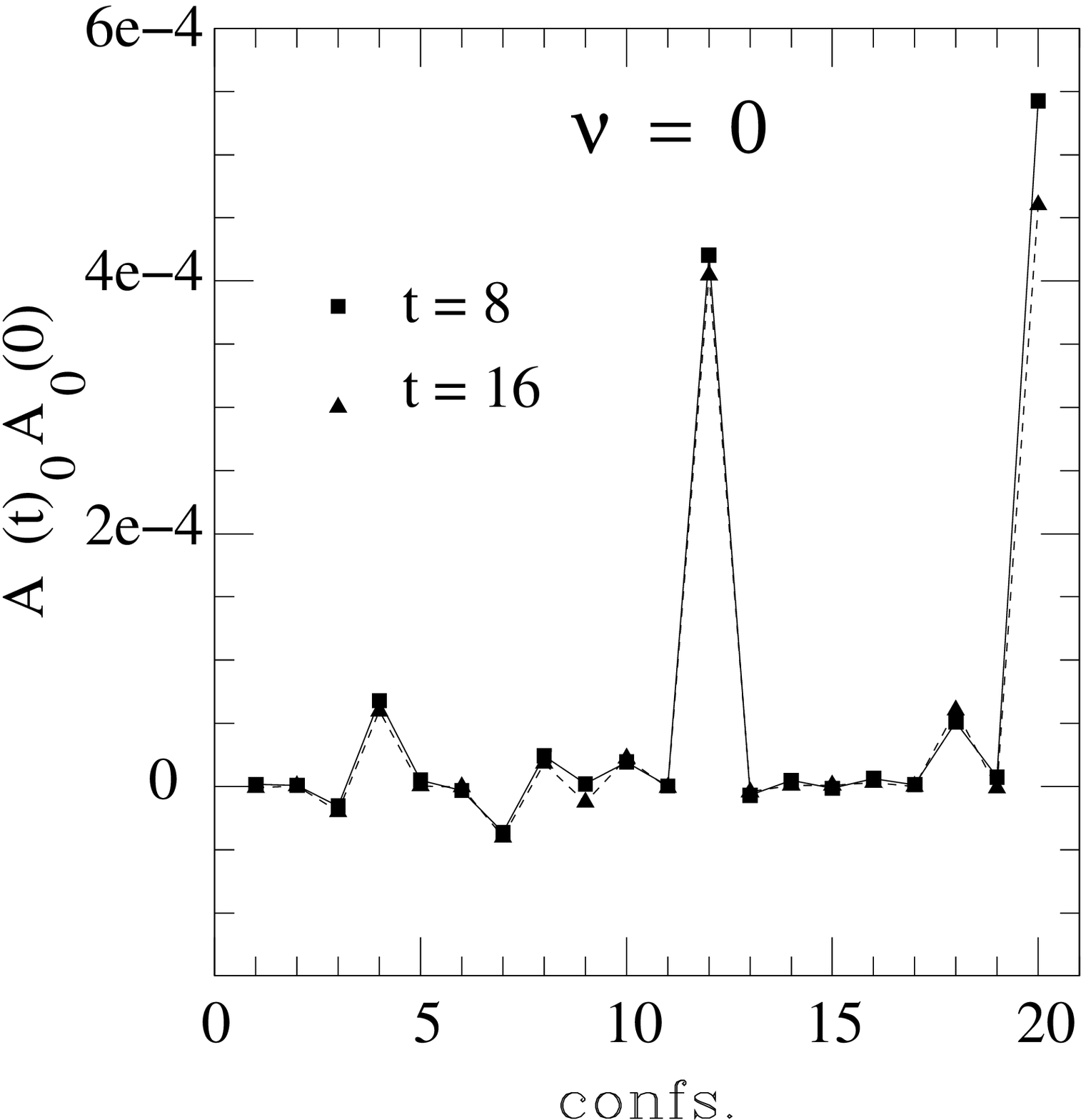}
\def\fpsangle{0}
\epsfxsize=40mm
\fpsbox{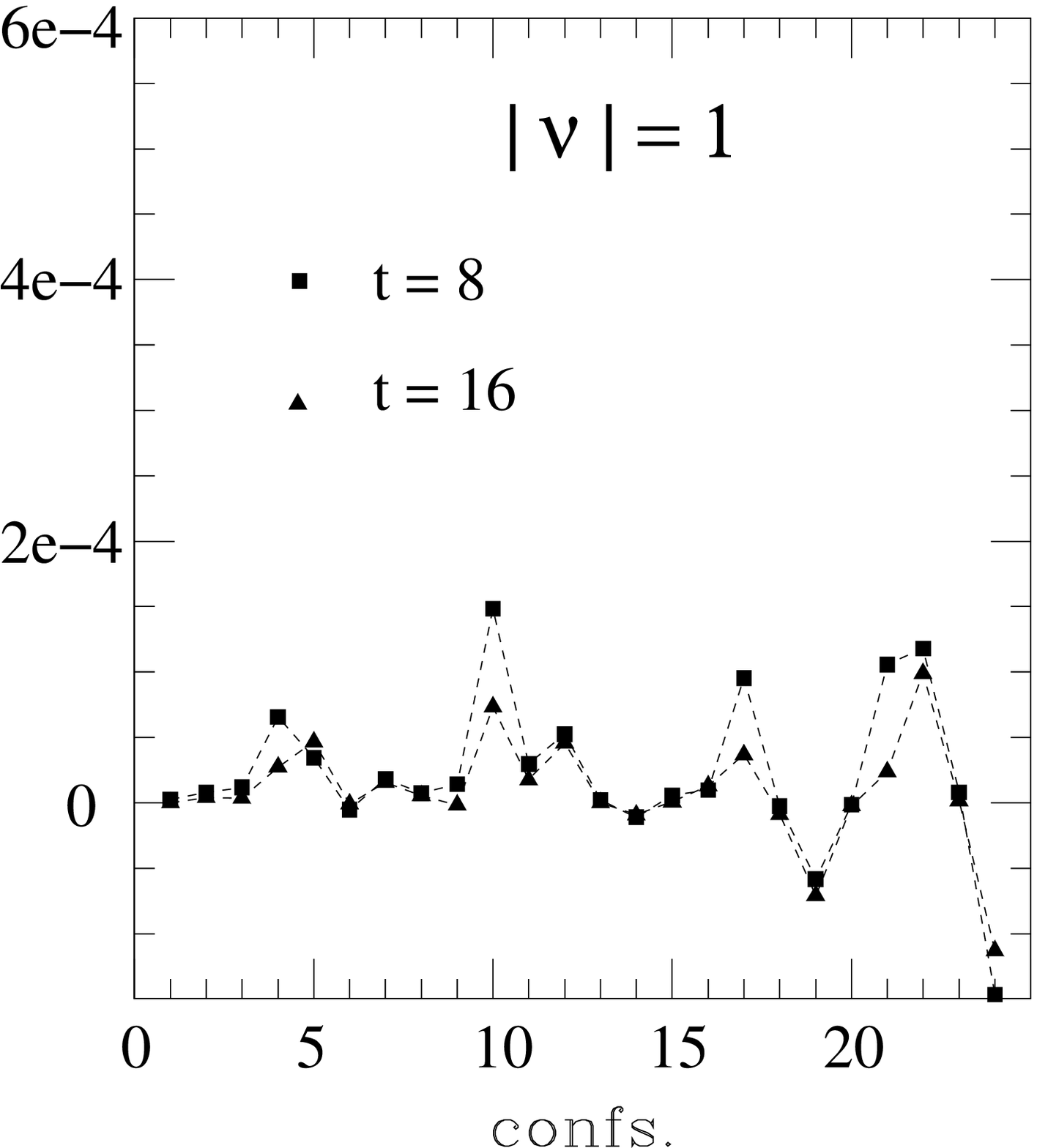}
\end{tabular}
\vspace*{-9mm}
\caption{\it{Histories of the axial correlation function 
$A_{0}(t) A_{0}(0)$ at $t=8$ and at $t=16$, with $m=21.3 ~{\rm MeV}$. 
For some $\nu =0$ configurations (on the left) we recognize
pronounced spikes. At $\vert \nu \vert =1$ (on the right)
the history is much smoother.}}
\label{history}
\vspace*{-7mm}
\end{figure}

\begin{figure}[hbt]
\vspace*{-3mm}
\hspace*{1cm}
\def\fpsangle{0}
\epsfxsize=47mm
\fpsbox{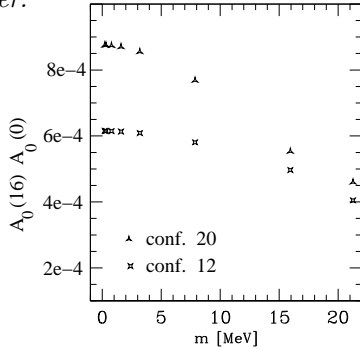}
\vspace*{-8mm}
\caption{\it{The height of the two spikes of $A_{0}(16) A_{0}(0)$
in the $\nu =0$ history of Fig.\ \ref{history},
at different quark masses $m$.}}
\label{spikes}
\vspace*{-8mm}
\end{figure}

We now address the {\em meson correlators} as physical observables,
from which one can try to extract $F_{\pi}$ and $\Sigma$.
The corresponding formulae for quenched $\chi$PT have been worked out
in Ref.\ \cite{corre}. For a first comparison to numerical
data we refer to Refs.\ \cite{TC}. The axial current
correlation function on an $L^3 \times T$ lattice,
\bea  \label{axial}
\vspace*{-1mm}
\hspace*{-7mm} && \la A_{\mu}(t) \, A_{\mu}(0) \ra_{\nu} = 
\frac{F_{\pi}^{2}}{T} 
+ 2 m \, \Sigma_{\vnu}(z) \, T \cdot
h_{1}(\tau ) \\
\hspace*{-7mm} &&
2 h_{1}(\tau ) = \tau^{2} - \tau + 1/6 
\ , \quad \tau = t/T \nn \\
\hspace*{-7mm} &&
\frac{\Sigma_{\nu}(z)}{\Sigma} = z \Big[ I_{\nu}(z) K_{\nu}(z) + 
I_{\nu +1}(z) K_{\nu -1}(z) \Big] + \frac{\nu}{z} \nn
\vspace*{-3mm}
\eea
is most suitable for our purpose, because it only involves
$F_{\pi}$ and $\Sigma$ as free parameters. The determination
of $F_{\pi}$ is relatively easy in the minimum at $\tau = 1/2$.
$\Sigma$ is related to the curvature and far more difficult
to extract, since the sensitivity of this curvature to a change
of $\Sigma$ is hardly visible over a wide range
\footnote{This situation is somehow
complementary to the RMT comparison of the EVs, where
$\Sigma$ is easier to determine \cite{BJS}.}.
However, we saw also here
that the physical volume must exceed the lower limit given before
in the context of RMT, otherwise one does not obtain the curvature
required by any acceptable value of $\Sigma$.
Once we are in the right regime, $\nu =0$ may look
like the simplest case from eq.\ (\ref{axial}), 
but from the numerical point of view this
is a nightmare. The problems in this sector are
related to the danger of very small EVs. Ref.\ \cite{HJL}
pointed out before that they may be a serious problem,
and our observations confirm this in a striking way.
As an example, we show histories for the measurement of
$\la A_{0}(t) \, A_{0}(0) \ra_{0}$ and
$\la A_{0}(t) \, A_{0}(0) \ra_{1}$ 
in Fig.\ \ref{history}. In the topologically neutral sector
there are strong spikes at those configurations which have
very small EVs. Their height is maximal at very small quark mass,
as Fig.\ \ref{spikes} illustrates.

\begin{figure}[hbt]
\hspace*{4mm}
\def\fpsangle{270}
\epsfxsize=48mm
\fpsbox{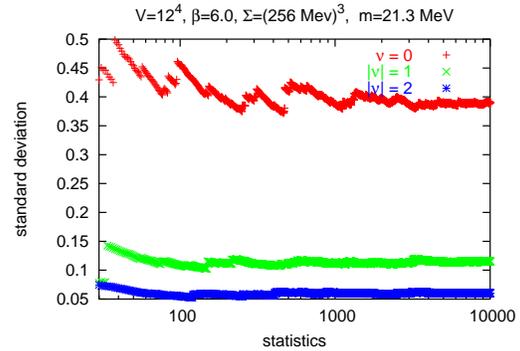}
\vspace*{-8mm}
\caption{\it{The standard deviation in the (fake) measurement
of $\Sigma_{\rm min}^{(\nu )}$, the contribution of the lowest 
non-zero EV to the scalar condensate.}}
\label{Smin}
\vspace*{-7mm}
\end{figure}

Obviously the occurrence of such spikes makes it very hard to
measure expectation values --- they stabilize only with a huge
statistics. To get an estimate for this effect,
we consider the contribution of the smallest (non-zero) EV alone
to the scalar condensate. This contribution reads
\be
\Sigma_{\rm min}^{(\nu )} = \frac{1}{V} \int_{0}^{\infty} dz \,
P_{\nu}(z) \frac{2m}{m^{2} + (z/\Sigma V)^{2}} \ ,
\ee
where we insert the probabilities $P_{\nu}(z)$ provided by
RMT. We now generated fake values for the lowest EV
with probability $P_{\nu}(z)$ and computed $\Sigma_{\rm min}^{(\nu )}$
from them. The result can only be trusted when the standard
deviation becomes practically constant.
Fig.\ \ref{Smin} shows the evolution of this standard deviation
as the statistics is enhanced. We see that the sector $\nu =0$
would require a tremendous statistics of $O(10^{4})$ configurations,
but the situation is clearly better for $\nu \neq 0$.

We conclude that there are several conditions for running
conclusive simulations with Ginsparg-Wilson fermions in the $\eps$-regime.
The volume should obey $V \gsim (1.2~{\rm fm})^{4}$, and
measurements have to be performed in a sector of
{\em non-trivial topology.}
The quark mass should be small for conceptual reasons,
but taking it too small causes technical problems in the
measurements.

\end{document}